\documentclass[]{aastex631}

\usepackage{xcolor}
\usepackage{soul}
\usepackage{CJK}

\shorttitle{An old brown dwarf in a hierarchical triple system with two white dwarfs}
\shortauthors{Rebassa-Mansergas et al.}
\graphicspath{{./}{figures/}}

\begin{document}

\title{\emph{Gaia}\,0007-1605: an old triple system with an inner
  brown dwarf-white dwarf binary and an outer white dwarf companion}

\correspondingauthor{Alberto Rebassa-Mansergas}
\email{alberto.rebassa@upc.edu}

\author[0000-0002-6153-7173]{Alberto Rebassa-Mansergas}
\affiliation{Departament de F\'{\i}sica, Universitat Polit\`ecnica de Catalunya, c/ Esteve Terrades, 5, 08860 Castelldefels, Spain}
\affiliation{Institut d'Estudis Espacials de Catalunya, Ed. Nexus-201, c/Gran Capit\`a 2-4, 08034 Barcelona, Spain}

\author[0000-0002-8808-4282]{Siyi Xu \begin{CJK*}{UTF8}{gbsn}(许\CJKfamily{bsmi}偲\CJKfamily{gbsn}艺\end{CJK*})}
\affiliation{Gemini
Observatory/NSF's NOIRLab, 670 N. A'ohoku Place, Hilo, Hawaii, 96720, USA}

\author[0000-0002-9090-9191]{Roberto Raddi}
\affiliation{Departament de F\'{\i}sica, Universitat Polit\`ecnica de Catalunya, c/ Esteve Terrades, 5, 08860 Castelldefels, Spain}

\author[0000-0001-7069-7403]{Anna F. Pala}
\affiliation{European Space Agency, European Space Astronomy Centre, Camino Bajo del Castillo s/n, 28692 Villanueva de la Ca\~nada, Madrid, Spain}

\author[0000-0003-1885-5130]{Enrique Solano}
\affiliation{Departmento de Astrof\'{\i}sica, Centro de Astrobiolog\'{\i}a (CSIC-INTA), ESAC Campus, Camino Bajo del Castillo s/n,\\
E-28692 Villanueva de la Ca\~nada, Madrid, Spain}
\affiliation{Spanish Virtual Observatory, E-28692 Villanueva de la Ca\~nada, Madrid, Spain}

\author[0000-0001-5777-5251]{Santiago Torres}
\affiliation{Departament de F\'{\i}sica, Universitat Polit\`ecnica de Catalunya, c/ Esteve Terrades, 5, 08860 Castelldefels, Spain}
\affiliation{Institut d'Estudis Espacials de Catalunya, Ed. Nexus-201, c/Gran Capit\`a 2-4, 08034 Barcelona, Spain}

\author[0000-0002-6985-9476]{Francisco Jim\'enez-Esteban}
\affiliation{Departmento de Astrof\'{\i}sica, Centro de Astrobiolog\'{\i}a (CSIC-INTA), ESAC Campus, Camino Bajo del Castillo s/n,\\
E-28692 Villanueva de la Ca\~nada, Madrid, Spain}
\affiliation{Spanish Virtual Observatory, E-28692 Villanueva de la Ca\~nada, Madrid, Spain}

\author[0000-0003-1793-200X]{Patricia Cruz}
\affiliation{Departmento de Astrof\'{\i}sica, Centro de Astrobiolog\'{\i}a (CSIC-INTA), ESAC Campus, Camino Bajo del Castillo s/n,\\
E-28692 Villanueva de la Ca\~nada, Madrid, Spain}
\affiliation{Spanish Virtual Observatory, E-28692 Villanueva de la Ca\~nada, Madrid, Spain}



\begin{abstract}
We  identify  \emph{Gaia}\,0007-1605\,A,C  as the  first  inner  brown
dwarf-white dwarf binary of a  hierarchical triple system in which the
outer  component is  another white  dwarf (\emph{Gaia}\,0007-1605\,B).
From  optical/near-infrared spectroscopy  obtained at  the Very  Large
Telescope  with  the  X-Shooter  instrument  and/or  from  \emph{Gaia}
photometry plus  SED fitting, we determine  the effective temperatures
and    masses   of    the   two    white   dwarfs    (12018$\pm$68\,K,
0.54$\pm$0.01\,M$_{\odot}$    for     \emph{Gaia}\,0007-1605\,A    and
4445$\pm$116\,K,             0.56$\pm$0.05\,M$_{\odot}$            for
\emph{Gaia}\,0007-1605\,B) and the effective  temperature of the brown
dwarf (1850$\pm$50\,K; corresponding to a spectral type L3$\pm$1).  By
analysing    the    available    {\em   TESS}    light    curves    of
\emph{Gaia}\,0007-1605\,A,C     we     detect      a     signal     at
1.0446$\pm$0.0015\,days  with  an  amplitude  of 6.25  ppt,  which  we
interpret as the orbital period  modulated from irradiation effects of
the  white dwarf  on  the brown  dwarf's surface.  This  drives us  to
speculate  that the  inner binary  evolved through  a common  envelope
phase in the  past. Using the outer white dwarf  as a cosmochronometer
and analysing the kinematic properties of the system, we conclude that
the triple system is about 10 Gyr old.
\end{abstract}



\section{Introduction} \label{sec:intro}

Triple star systems are common, with fractions of $\simeq$10\% for F-G
stars \citep{Tokovinin2014} and increasing  up to $\simeq$50\% for O-B
stars    \citep{Sana2014}.   The    majority    of   triple    systems
($\simeq$70-80\%)  are   believed  to  interact  during   their  lives
\citep{Toonen2020}.  Mass transfer  episodes may  take place  from one
star to another,  in which case a common envelope  phase ensues if the
process is dynamically unstable \citep{Webbink2008}. Moreover -- since
the stellar  components are  subject to their  individual evolutionary
paths thus changing their masses and radii if they have time to evolve
out of the main sequence -- three-body dynamics are expected to modify
the  systems, leading  to  tidal interactions,  collisions or  mergers
\citep{Antonini2017}.   In  this  context,  a clear  example  are  Von
Ziepel-Lidov-Kozai  oscillations  in  hierarchical triples,  with  the
outer component inducing eccentricity oscillations in the inner binary
\citep{Naoz2016}.   Tides   from  the  Galactic  potential   can  also
influence   the    orbits   of    the   stars   in    triple   systems
\citep{Grishin+Perets2021}.  These  mechanisms   may  lead  to  exotic
outcomes  such as  blue stragglers  \citep{Perets+Fabrycky2009}, black
hole \citep{Antonini2017} or neutron star mergers \citep{Liu+Lai2018},
type  Ia   supernovae  \citep{Katz+Dong2012}   or  gamma   ray  bursts
\citep{Thompson2011}.

\begin{deluxetable*}{ccccccccc}
\tablenum{1} \tablecaption{$Gaia$ EDR3 parameters \citep{GaiaEDR3} for
  the inner unresolved binary (\emph{Gaia}\,0007-1605\,A,C) and the
  outer white dwarf (\emph{Gaia}\,0007-1605\,B).
  \label{tab:info}}
\tablewidth{0pt}
\tablehead{
\colhead{Object} &  \colhead{RA} &
\colhead{DEC} & \colhead{$G$} & \colhead{$G_\mathrm{BP}$} & \colhead{$G_\mathrm{GRP}$} & \colhead{pm$_\mathrm{RA}$} & \colhead{pm$_\mathrm{DEC}$} & \colhead{parallax}  \\
\colhead{} &  \colhead{(deg)} & \colhead{(deg)} & \colhead{(mag)} & \colhead{(mag)} & \colhead{(mag)} & \colhead{(mas/yr)} & \colhead{(mas/yr)} & \colhead{(mas)}\\
}
\decimalcolnumbers
\startdata
\emph{Gaia}\,0007-1605\,A,C & 1.895921 & -16.09240 & 16.152$\pm$0.003 & 16.162$\pm$0.004 & 16.157$\pm$0.008 & 179.23$\pm$0.05 & -64.40$\pm$0.04 & 12.34$\pm$0.04 \\
\emph{Gaia}\,0007-1605\,B & 1.893281 & -16.08726 & 19.942$\pm$0.005 & 20.59$\pm$0.08 & 19.24$\pm$0.05 & 176.36$\pm$0.61 & -64.11$\pm$0.41 & 12.53$\pm$0.55 \\
\enddata
\end{deluxetable*}

Given the  variety of possible evolutionary  scenarios, triple systems
come     in    different     flavours:    main     sequence    triples
\citep{Kervella2017},     main     sequence/brown    dwarf     triples
\citep{Faherty2011},  brown  dwarf triples  \citep{Triaud2020},  giant
star/brown dwarf triples  \citep{Lillo2021}, main sequence/white dwarf
triples   \citep{Toonen2017},   main   sequence/black   hole   triples
\citep{Rivinius2020}, white dwarf  triples \citep{Perpinya2019}, white
dwarf/neutron   star    triples   \citep{Ransom2014},    etc.   Hence,
observational studies  of triple  systems are  not only  important for
constraining our understanding of  multiple stellar evolution but also
to elucidate the origin of such exotic objects.

In this  work we identify the  first trinary star formed  by two white
dwarfs and a brown dwarf. The structure of the system is hierarchical,
with   an   inner   white    dwarf-brown   dwarf   binary   (hereafter
\emph{Gaia}\,0007-1605\,A,C)  and   a  wider  white   dwarf  companion
(hereafter \emph{Gaia}\,0007-1605\,B). Using the  outer white dwarf as
a cosmochronometer, we find this unique  system to be nearly as old as
the age of the Galactic disk, thus  implying the brown dwarf to be one
of the oldest known of its kind.

\section{Identification of \emph{Gaia}\,0007-1605}
\label{s-ident}

\emph{Gaia}\,0007-1605\,A,C  (ID 2416481783371550976),  also known  as
EGGR509, was identified as an infrared-excess white dwarf candidate by
\citet{Rebassa2019},  who analysed  the spectral  energy distributions
(SED) of  3733 \emph{Gaia} white  dwarfs within 100\,pc  with reliable
infrared  photometry and  $G_\mathrm{BP}-G_\mathrm{RP}  <0.8$ mag.  We
obtained a  spectrum (430\,s exposure)  for this object with  the Very
Large Telescope at  Cerro Paranal (Chile) equipped  with the X-Shooter
instrument and the  1" slits on the night of  December 3rd 2020, which
revealed the infrared  excess to arise due to the  presence of a brown
dwarf companion\footnote{The spectra were reduced/calibrated using the
  X-Shooter pipeline  version 2.11.5.  Telluric removal  was performed
  using the Molecfit software.} (see Figure\,\ref{fig:fitspec}).

The  measured \emph{Gaia}  parallax of  \emph{Gaia}\,0007-1605\,A,C is
12.34$\pm$0.04  mas (Table\,\ref{tab:info})  and the  inverse parallax
yields   a  distance   of  81.1$\pm$0.1   pc.  \citet{Badry2021}   and
\citet{Rebassa2021a} report this object as a common proper motion pair
member with a second white  dwarf (ID 2416481779075909376) at the same
distance  and proper  motions  (Table\,\ref{tab:info}). The  projected
separation  between \emph{Gaia}\,0007-1605\,A,C  and  the outer  white
dwarf  \emph{Gaia}\,0007-1605\,B is  1673.11 AU  \citep{Badry2021} and
the radial separation is 1.2 pc \citep{Torres2022}.

\begin{deluxetable*}{ccccccccc}
\tablenum{2} \tablecaption{Stellar parameters and  ages for the triple
  system studied in this work. The method used to derive the effective
  temperatures is also indicated.
  \label{tab:param}}
\tablewidth{0pt}
\tablehead{
\colhead{Object} & \colhead{Type} & \colhead{$T_\mathrm{eff}$} & \colhead{$\log$\,g} & \colhead{mass} & \colhead{t$_\mathrm{cool}$} & \colhead{Total Age} & \colhead{$P_\mathrm{orb}$} & Method \\
\colhead{} & \colhead{}  & \colhead{(K)} & \colhead{(dex)} & \colhead{(M$_{\odot}$)} & \colhead{(Gyr)} & \colhead{(Gyr)} & \colhead{(days)} & \\
}
\decimalcolnumbers
\startdata
\emph{Gaia}\,0007-1605\,A & DA  & 12018$\pm$68 & 7.87$\pm$0.02 & 0.54$\pm$0.01 & 0.360$\pm$0.002 & $\simeq$10* & 1.0446$\pm$0.0015 & Phot./Spect. \\
\emph{Gaia}\,0007-1605\,C & L3  & 1850$\pm$50  & -              & 0.07*           &    -          & $\simeq$10* & 1.0446$\pm$0.0015 & Spect. \\
\emph{Gaia}\,0007-1605\,B & pure-H DC* & 4445$\pm$116 & 7.96$\pm$0.05  & 0.56$\pm$0.05     & 8.2$\pm$0.2 & $\simeq$10* & - & Phot. \\
\enddata
\tablecomments{The values and parameters indicated by an * are
  assumptions and require confirmation.}
\end{deluxetable*}

\section{Stellar parameters of the three components}
\label{s-param}

\begin{figure*}
    \centering
    \includegraphics[angle=-90,width=0.8\textwidth]{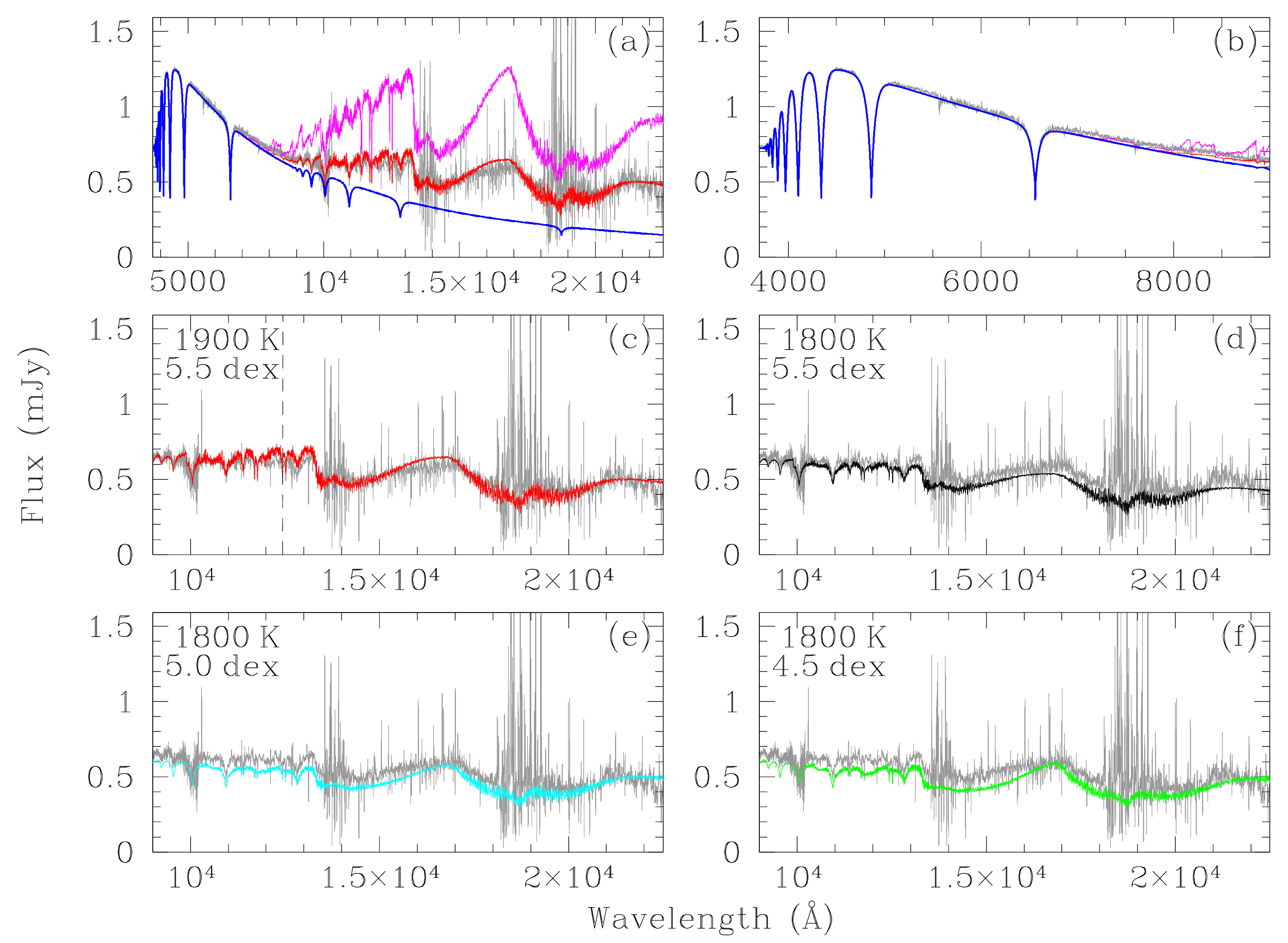}
    \caption{Panel   (a):   the   observed   X-Shooter   spectrum   of
      \emph{Gaia}\,0007-1605\,A,C  (grey),  the best-fit  white  dwarf
      model  (blue) and  its combination  with a  2300\,K and  5.0 dex
      CIFIST model (magenta; corresponding to  a star of spectral type
      M9.5) and a 1900\,K and 5.5  dex CIFIST model, i.e. the best-fit
      to the observed spectrum. An M9.5 companion can clearly be ruled
      out, which  confirms the  near-infrared excess  to arise  from a
      brown   dwarf.   The   structure  at   $\simeq$14.000\AA\,   and
      $\simeq$19.000\AA\, is related to  the residuals of the telluric
      correction. Panel (b): the same but for the optical range of the
      spectrum, where  the white dwarf  dominates the SED.  Panel (c):
      the  combined  best-fit  to   the  near-infrared  range  of  the
      spectrum. The  effective temperature and surface  gravity of the
      brown dwarf are indicated in the  top left corner and the dashed
      vertical       line      indicates       the      K$_\mathrm{I}$
      12\,432.27/12\,522.14\,\AA\  absorption  doublet from  which  we
      measure  the  radial velocity.  Panels  (d),  (e) and  (f):  for
      comparison  we  show the  second-best  (d),  third-best (e)  and
      fourth-best  (f)  CIFIST  models superimposed  to  the  observed
      near-infrared spectrum.  The effective temperatures  and surface
      gravities are also indicated.}
    \label{fig:fitspec}
\end{figure*}

We  derived  three independent  values  of  effective temperature  and
surface gravity  for the  inner white  dwarf \emph{Gaia}\,0007-1605\,A
from:  (1)   the  available  \emph{Gaia}  EDR3   $G_\mathrm{abs}$  and
$G_\mathrm{BP}-G_\mathrm{RP}$ colours  following the  method described
in  \citet{Rebassa2021a}\footnote{Note that  the  brown  dwarf in  the
  inner binary is completely out-shined in  the optical by the flux of
  the white  dwarf.  Only  for wavelengths  larger than  8000\AA\, the
  average   excess  flux   contribution  from   the  brown   dwarf  is
  $\simeq$5\%. Therefore,  the reported \emph{Gaia} magnitudes  can be
  safely  considered as  those  arising from  the  white dwarf  only}.
($T_\mathrm{eff}$=11814$\pm$142\,K  and  $\log g$=7.86$\pm$0.02  dex);
(2)  fitting  the  entire  SED using  VOSA  (Virtual  Observatory  SED
Analyser)\footnote{The   VOSA  documentation   can  be   accessed  via
  \url{http://svo2.cab.inta-csic.es/theory/vosa/help/star/intro}.}
following  \citet{Jimenez2018} ($T_\mathrm{eff}$=12000$\pm$125\,K  and
$\log g$=7.84$\pm$0.03 dex;  the passbands used in the  fit were GALEX
NUV/FUV,    APASS   BgVi,    Pan-STARRS   grizy,    \emph{Gaia}   EDR3
$G/G_\mathrm{BP}/G_\mathrm{RP}$, DENIS  IJ, 2MASS JHK and  WISE w1w2);
(3) fitting  the optical X-Shooter  spectrum with the updated  grid of
white dwarf  model atmosphere  spectra of  \citet{Koester2010}, taking
into    account   the    3D   corrections    by   \citet{Tremblay2013}
($T_\mathrm{eff}$=12131$\pm$100\,K and $\log g$=7.97$\pm$0.04 dex; the
best-fit     white      dwarf     model     is      illustrated     in
Figure\,\ref{fig:fitspec}).      The      weighted     mean     yields
$T_\mathrm{eff}$=12018$\pm$68\,K and $\log g$=7.87$\pm$0.02 dex, which
are    our     adopted    parameters    for    this     white    dwarf
(Table\,\ref{tab:param}).      For    the     outer    white     dwarf
\emph{Gaia}\,0007-1605\,C  there is  no available  spectrum and,  as a
consequence, we  derived two  independent set  of parameters  from the
\emph{Gaia} $G_\mathrm{abs}$ and $G_\mathrm{BP}-G_\mathrm{RP}$ colours
($T_\mathrm{eff}$=4121$\pm$304\,K  and $\log  g$=7.8$\pm$0.2 dex)  and
VOSA ($T_\mathrm{eff}$=4500$\pm$125\,K  and $\log  g$=8.0$\pm$0.1 dex;
the  passbands  used in  the  fit  were  Pan-STARRS grizy,  Gaia  EDR3
$G/G_\mathrm{BP}/G_\mathrm{RP}$  and  WISE  w1).  Hence,  our  adopted
values are $T_\mathrm{eff}$=4445$\pm$116\,K and $\log g$=7.96$\pm$0.05
dex  for  the outer  white  dwarf  (Table\,\ref{tab:param}).  We  then
interpolated the adopted effective  temperatures and surface gravities
of  the white  dwarfs  in  the cooling  sequences  of  La Plata  group
\citep{Althaus2015,  Camisassa2016,  Camisassa2019}  to  obtain  their
masses (Table\,\ref{tab:param}).

For the  above calculations we  assumed the  two white dwarfs  to have
hydrogen-rich atmospheres. This is confirmed for the inner white dwarf
(Figure\,\ref{fig:fitspec})   but  it   has   not   been  tested   for
\emph{Gaia}\,0007-1605\,C.   For   this  object,   \citet{Gentile2021}
derives   4500$\pm$400\,K   (assuming   either  a   hydrogen-rich   or
helium-rich  atmosphere), in  agreement with  our value.  At such  low
effective temperature a spectrum would be featureless (DC type), hence
giving no indications about the  atmospheric composition of this white
dwarf.

In  order  to  derive  the  brown dwarf  parameters,  we  performed  a
composite  spectral fit.  We created  a set  of composite  models that
combine  the  white dwarf  best-fit  model  with  a grid  of  BT-Settl
(CIFIST; \citealt{Allard2013},  only at  solar metallicity\footnote{We
  also  used low-metallicity  NextGen models  to evaluate  whether the
  brown dwarf  could be an ultracool  subdwarf star but found  no good
  fits in any of these cases.}) low-mass and brown dwarf model spectra
and obtained a $\chi^2$ value  between the observed X-Shooter spectrum
and each  combined model.  The CIFIST grid  contained 42  spectra with
effective temperatures  in the range  1300-2600\,K in steps  of 100\,K
and surface gravities 4.5-5.5 dex in steps of 0.5 dex. All models were
scaled  to the  distance  of 81.1  pc  by assuming  a  radius from  an
effective  temperature-radius relation  for low-mass  stars and  brown
dwarfs  \citep{Pecaut+Mamajek2013,   Baraffe2015}  and   an  effective
temperature-radius relation for white dwarfs from La Plata tracks. The
CIFIST model associated to the lowest $\chi^2$ ($\chi^2_\mathrm{min}$)
was that of an effective temperature  of 1900\,K and a surface gravity
of 5.5  dex (Figure\,\ref{fig:fitspec}).  Considering all models  at a
distance of  less than  1$\sigma$ from $\chi^2_\mathrm{min}$  as valid
solutions resulted in temperatures in  the range 1800-1900\,K with all
possible    surface    gravities    (4.5,    5.0    and    5.5    dex;
Figure\,\ref{fig:fitspec}). Therefore, we cannot constrain the surface
gravity of  the brown  dwarf but  we can  estimate its  temperature as
1850$\pm$50\,K.  This  corresponds  to  a spectral  type  of  L3$\pm$1
\citep{Nakajima2004}.

\begin{figure}
    \centering
    \includegraphics[width=0.4\columnwidth]{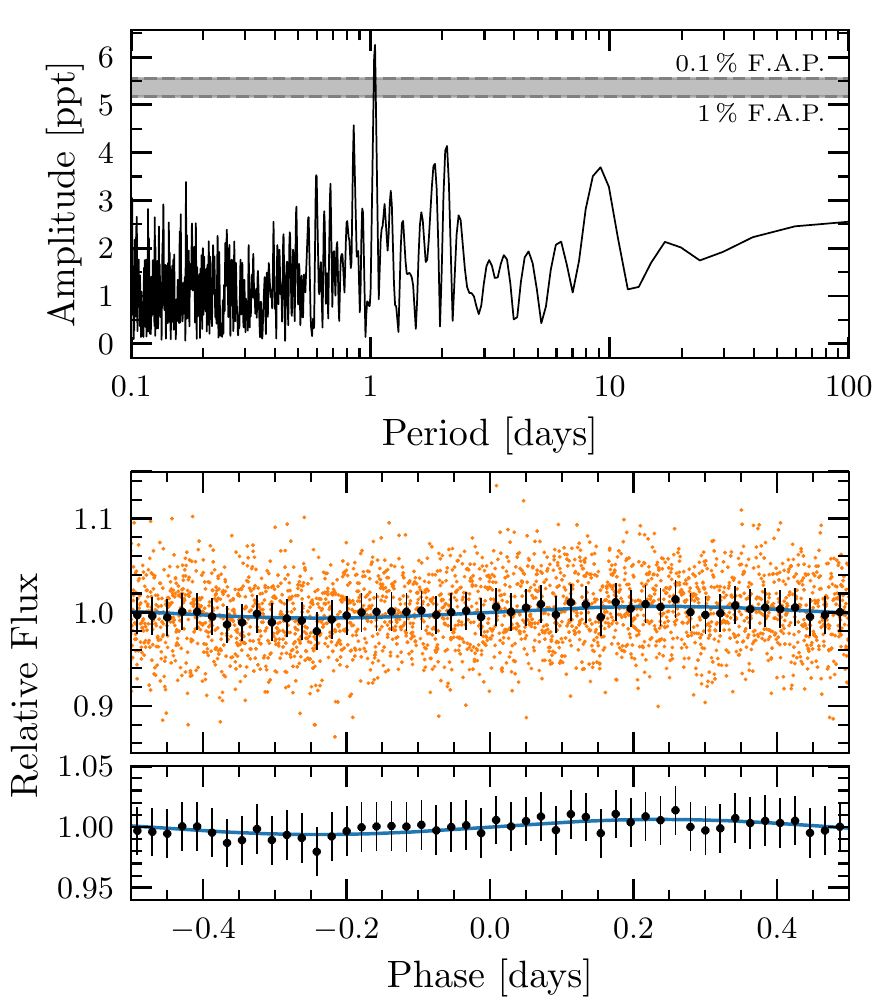}
    \caption{Top  panel: Lomb-Scargle  periodogram of  the {\em  TESS}
      light-curve.   The strongest  signal at  1.0446$\pm$0.0015\,days
      exceeds the  1 and 0.1\,\% false-alarm-probability  (F.  A.  P.)
      thresholds. Middle panel: folded  light-curve, obtained from the
      1.0446\,days  period.    The  black   symbols  are   the  binned
      light-curve at 30-min intervals and the error bars represent the
      1-$\sigma$  scatter  in  each  bin. The  blue  sinusoidal  curve
      corresponds  to  the  detected periodic  signal.  Bottom  panel:
      zoomed panel on the binned light-curve.}
    \label{fig:tess}
\end{figure}

\section{Observational clues on the past evolution of the inner binary}
\label{s-evol}

\emph{Gaia}\,0007-1605\,A,C  is   unresolved  by   \emph{Gaia},  which
probably implies it evolved through  a common envelope phase. In order
to  test  this  hypothesis   we  searched  for  available  time-series
photometry  from the  Mikulski Archive  for Space  Telescopes, finding
that this  object was  observed by the  {\em Transit  Exoplanet Survey
  Satellite}   \citep[{\em    TESS};][]{tess}   in   Sector    29   as
TIC\,289593425,  between 2020  August 26  and 2020  September 21.   We
analysed  the 20  second, 2-  and 10-min  cadence data  using standard
tools   of   the  \verb|ligtkurve|   software   finding   a  peak   at
1.0446$\pm$0.0015\,days in the Lomb-Scargle  periodogram of the 10-min
cadence light-curve (see Figure\,\ref{fig:tess}, top panel). This peak
corresponds to a  signal amplitude of 6.25 ppt that  is just above the
0.1\,\%  false-alarm-probability   limit  that   we  computed   as  in
\citet{hermes2017},  meaning  that  the detection  has  a  probability
higher   than    99.9\%   of   not-being   caused    by   noise.    In
Figure\,\ref{fig:tess} (middle and bottom  panels), we show the folded
light-curve  for  this  period.   The light-curve  is  clearly  noisy,
however, by  binning it at every  30-min we are able  to highlight the
periodic variability detected  by {\em TESS}. We  interpret the signal
at 1.0446-day  period as  the irradiation  of the  white dwarf  on the
brown dwarf's surface that becomes visible to us at every orbit.

\begin{figure}
    \centering
    \includegraphics[angle=-90,width=0.4\columnwidth]{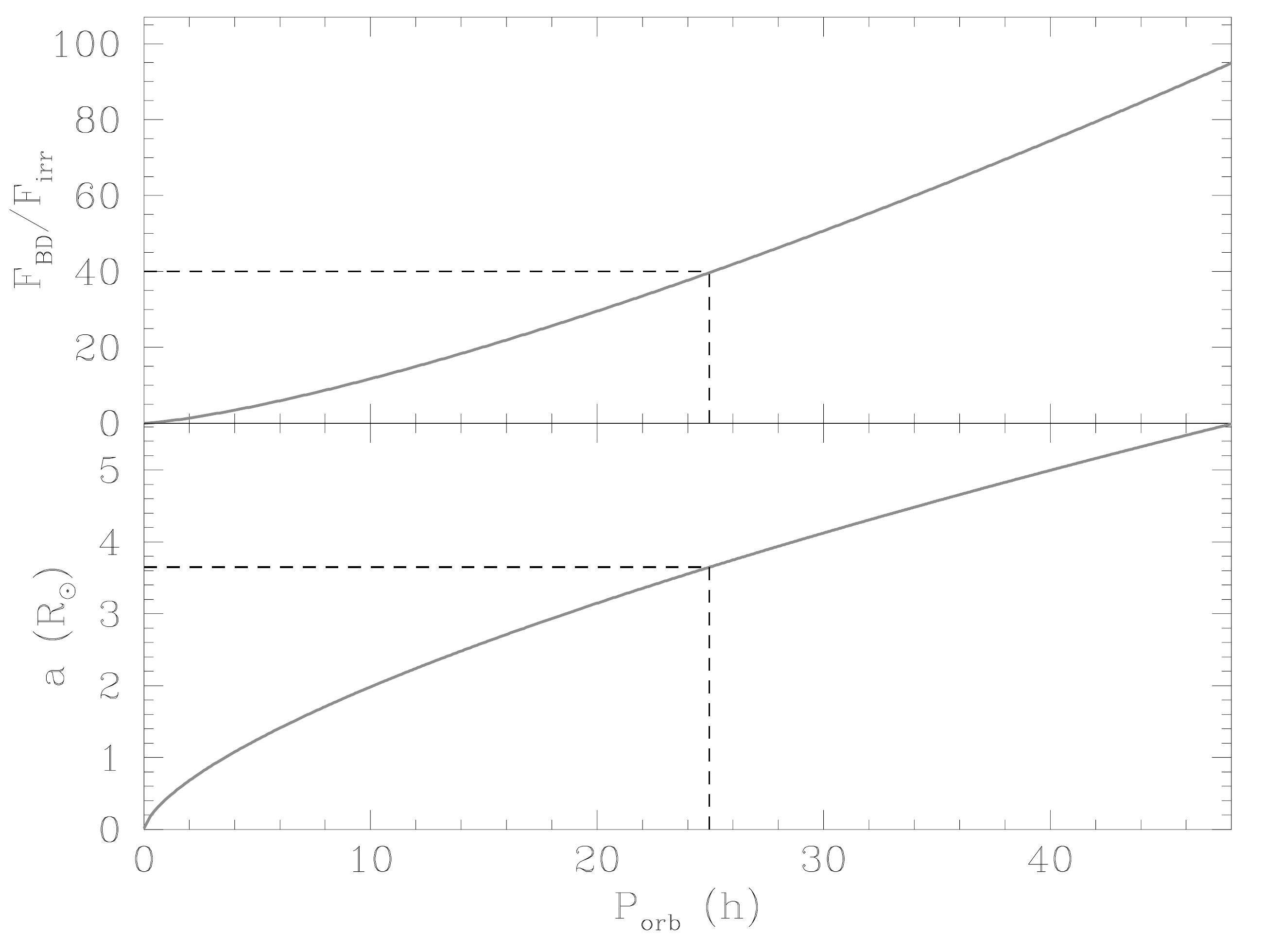}
    \caption{Top  panel:  the ratio  between  the  brown dwarf's  flux
      ($F_\mathrm{BD}$) and  the irradiated flux from  the white dwarf
      on the brown dwarf's surface ($F_\mathrm{irr}$) as a function of
      orbital  period.   Bottom panel:  the  orbital  separation as  a
      function of orbital period. The dashed black lines represent the
      values  for 1.0446  days, which  is  assumed to  be the  orbital
      period of \emph{Gaia}\,0007-1605\,A,C.}
    \label{fig:fluxr}
\end{figure}

The ratio  between the  brown dwarf's  flux ($F_\mathrm{BD}$)  and the
irradiated  flux from  the white  dwarf on  the brown  dwarf's surface
($F_\mathrm{irr}$) can be expressed as \citep{Rebassa2013}:
\begin{equation}
\frac{F_\mathrm{BD}}{F_\mathrm{irr}} = \left(\frac{T_\mathrm{eff,BD}}{T_\mathrm{eff,WD}}\right)^4 \times \left(\frac{a}{R_\mathrm{WD}}\right)^2,
\end{equation}

\noindent
where $T_\mathrm{eff,BD}$ and $T_\mathrm{eff,WD}$  are the brown dwarf
and white  dwarf effective temperatures, $R_\mathrm{WD}$  is the white
dwarf  radius  and $a$  is  the  orbital  separation.  All  are  known
parameters (Section\,\ref{s-param};  note that the white  dwarf radius
can  be  obtained  from  the  surface gravity  and  mass  reported  in
Table\,\ref{tab:param})  except $a$,  which we  derived from  Kepler's
third law assuming 1.0446 days as  the orbital period of the binary as
well as via adopting a white dwarf and a brown dwarf mass. The mass of
the white dwarf is known, but  the brown dwarf's mass is not. However,
following      the      spectral       type-mass      relation      of
\citet{Pecaut+Mamajek2013}\footnote{see                           also
  \url{http://www.pas.rochester.edu/~emamajek/EEM_dwarf_UBVIJHK_colors_Teff.dat}.}
we estimated  the mass to  be 0.07\,M$_{\odot}$. Given that  the white
dwarf's mass is  considerably higher than the brown  dwarf's mass, the
orbital separation  does not  heavily depend  on this  assumption.  In
fact,  using   masses  between   0.05  and  0.1   M$_{\odot}$  yielded
practically identical  orbital separations. The top  and bottom panels
of Figure\,\ref{fig:fluxr} show the  flux ratio and orbital separation
as a function of the orbital period, respectively, which correspond to
$F_\mathrm{BD}/F_\mathrm{irr}$=40  and   $a$=3.65\,R$_{\odot}$  for  a
period of 1.0446 days.

Based on the above calculations,  the fraction of irradiated flux from
the white dwarf  on the brown dwarf's surface is  $\simeq$2.5\% of the
total  flux. This  percentage should  be  lower if  one considers  the
6\,000-10\,000\AA\, range sampled by  the \emph{TESS} bandpass filter,
since the  white dwarf's  flux peaks  in the  ultraviolet. This  is in
agreement with the amplitude modulation of $\simeq$0.5\% in the folded
light-curve (middle and bottom panels of Figure\,\ref{fig:tess}).

For completeness,  we also  derived the radial  velocity of  the brown
dwarf   via  fitting   the   $K_\mathrm{I}$   absorption  doublet   at
12\,432.27/12\,522.14\,\AA\, from the X-Shooter spectrum with a double
Gaussian  profile  of  fixed  separation.  To that  end  we  used  the
\verb|MOLLY|   software,   resulting    in   42$\pm$11   km/s,   which
unfortunately did  not provide much  information since we do  not know
the orbital inclination  of the inner binary nor the  orbital phase at
which the spectrum  was taken. For the white  dwarf, radial velocities
of  13.1$\pm$2.5  and  15.5$\pm$1.3  km/s were  measured  by  the  SPY
high-resolution survey \citep{Napiwotzki2020}.  This implies a minimum
orbital inclination of  the inner binary towards the line  of sight of
$\ga$45 degrees.

We conclude that \emph{Gaia}\,0007-1605\,A,C  likely evolved through a
common   envelope   phase  and   has   now   an  orbital   period   of
1.0446$\pm$0.0015\,days.

\section{The age of the system}
\label{s-age}

Given that two  white dwarfs are members of the  triple system studied
in this work, we can independently derive two total ages by summing up
the  white  dwarf  cooling  ages   to  the  main  sequence  progenitor
lifetimes.   To  that  end,  we  follow  the  procedure  described  in
\citet{Rebassa2021a}, which considers the La Plata evolutionary tracks
taking into account the full evolutionary path of the white dwarf from
the  zero-age main  sequence.   In  these sequences  the  mass of  the
hydrogen  envelope  is  not   fixed  but  obtained  from  evolutionary
calculations.

The  white dwarf  in the  inner binary  has likely  evolved through  a
common envelope phase. As consequence,  we can only derive its cooling
age (Table\,\ref{tab:param}), since  applying an initial-to-final mass
relation to derive the progenitor lifetime would not be adequate.  The
outer  white dwarf  is  located  1.2 pc  away  from  the inner  binary
(Section\,\ref{s-ident}).  Hence,  it is  expected that it  evolved in
isolation.   Taking  into account  the  uncertainties  in the  derived
effective temperature and  surface gravity, we obtained  a cooling age
of $8.2\pm0.2$\,Gyr. Therefore, 8\,Gyr is the 1$\sigma$ lower limit to
the age  of the system  since we need  to add the  progenitor lifetime
(which we cannot obtain because  the initial-to-final mass relation is
not well defined for white dwarf masses under $\la$0.55\,M$_{\odot}$).

The  random forest  analysis of  the 100\,pc  white dwarf  \emph{Gaia}
population by \citet{Torres2019} indicates that the white dwarf in the
inner binary is  a thin disk candidate, whereas the  outer white dwarf
is  a thick  disk  candidate.  This is  mainly  due  to the  different
temperatures   (hence    cooling   ages)    of   the    white   dwarfs
(Table\,\ref{tab:param}).  Additionally, we  performed Galactic  orbit
integration for the inner binary and the distant companion by adopting
either   a  null   value  or   the  brown   dwarf's  radial   velocity
(Section\,\ref{s-evol}) and  using the five  \emph{Gaia}'s astrometric
parameters and  correlations, as we  followed the methods  outlined by
\citet{Raddi2021}. In both cases the  orbital eccentricity and the $Z$
component of the  angular momentum of the two white  dwarfs are in the
region  of $e  = 0.16$  an  $L_z =  1700$\,kpc\,km/s, that  is at  the
separation between thin and thick  disk orbits. This supports the idea
of the  triple system being rather  old, as expected from  the cooling
age we  obtained for the  outer white dwarf. If  we set the  total age
limit to  $\simeq$10\,Gyr, which  is the  mean age  of the  thick disk
\citep{Sharma2019}, the  stellar parameters for the  outer white dwarf
that are consistent with this value within $\pm$0.5 Gyr errors are the
following:    $\simeq$4400\,K   for    the   effective    temperature,
$\simeq$8.0\,dex  of surface  gravity and  $\simeq$0.6\,M$_{\odot}$ of
mass.  These   values  are   in  agreement   with  the   estimates  of
4500$\pm$400\,K,   8.0$\pm$0.3\,dex    and   0.6$\pm$0.20\,M$_{\odot}$
obtained by \citealt{Gentile2021} for this white dwarf assuming either
hydrogen- or helium-rich atmospheres. We thus conclude that the age of
the triple system studied in this work is around 10\,Gyr.

\section{Conclusions}

We  have  identified and  analysed  a  remarkable hierarchical  triple
system formed by an inner brown  dwarf-white dwarf binary and an outer
white dwarf companion, the first of its kind. The most likely scenario
is that \emph{Gaia}\,0007-1605\,A and C were relatively close binaries
during  the  main  sequence  stage  of  \emph{Gaia}\,0007-1605\,A  and
evolved     through     a     common     envelope     phase,     while
\emph{Gaia}\,0007-1605\,B evolved  like a single  star. Alternatively,
it  is possible  that \emph{Gaia}\,0007-1605\,C  was scattered  to the
current position after \emph{Gaia}\,0007-1605\,A became a white dwarf,
through           different           dynamical           interactions
\citep[e.g.][]{OConnor2021}.   Additional   modeling  is   needed   to
understand the evolutionary history of this interesting system.

Using the  outer white dwarf  as a cosmochronometer and  analysing the
kinematic properties of  the system, we find that the  trinary star is
very old and has an age of  $\simeq$10 Gyr. Being nearly as old as the
disk of the  Galaxy, the brown dwarf  \emph{Gaia}\,0007-1605\,C is not
expected to be magnetically active  nor rapidly rotating, unless it is
tidally locked by the white  dwarf. Therefore, it could potentially be
used as a benchmark for testing theoretical models of brown dwarfs. In
particular,  future  observations  with  the  {\em  James  Webb  Space
  Telescope} can  help in  improving the  measurements of  the stellar
parameters of this object.


\begin{acknowledgements}
ARM  acknowledges   support  from   Grant  RYC-2016-20254   funded  by
MCIN/AEI/10.13039/501100011033 and by ESF Investing in your future. SX
is supported  by the  international Gemini  Observatory, a  program of
NSF's NOIRLab, which is managed by the Association of Universities for
Research in  Astronomy (AURA) under  a cooperative agreement  with the
National Science  Foundation, on behalf  of the Gemini  partnership of
Argentina,  Brazil, Canada,  Chile,  the Republic  of  Korea, and  the
United  States   of  America.  RR   has  received  funding   from  the
postdoctoral fellowship  programme Beatriu  de Pin\'os, funded  by the
Secretary of  Universities and Research (Government  of Catalonia) and
by  the Horizon  2020  programme  of research  and  innovation of  the
European Union under the  Maria Sk\l{}odowska-Curie grant agreement No
801370.  ST and  ARM acknowledge  support  from the  MINECO under  the
PID2020-117252GB-I00 grant. F.J.E. acknowledges financial support from
the Spanish MINECO/FEDER through the  grant MDM-2017-0737 at Centro de
Astrobiolog\'{i}a  (CSIC-INTA),  Unidad  de  Excelencia  Mar\'{i}a  de
Maeztu.  PC  acknowledges financial  support  from  the Government  of
Comunidad  Aut\'onoma  de  Madrid   (Spain),  via  postdoctoral  grant
Atracci\'on de  Talento Investigador 2019-T2/TIC-14760.  This research
has    made     use    of    the    Spanish     Virtual    Observatory
(\url{http://svo.cab.inta-csic.es})   supported  from   Ministerio  de
Ciencia e  Innovaci\'on through grant PID2020-112949GB-I00.  This work
has made use of data from the European Space Agency (ESA) mission {\it
  Gaia} (\url{https://www.cosmos.esa.int/gaia}), processed by the {\it
  Gaia}    Data   Processing    and    Analysis   Consortium    (DPAC,
\url{https://www.cosmos.esa.int/web/gaia/dpac/consortium}).    Funding
for the DPAC has been provided by national institutions, in particular
the  institutions   participating  in  the  {\it   Gaia}  Multilateral
Agreement.

Based  on  observations collected  at  the  European Organisation  for
Astronomical   Research  in   the   Southern   Hemisphere  under   ESO
programme(s) 106.213V.001 and 0106.D-0386(A).

\end{acknowledgements}

%

\vspace{5mm}
\facilities{VLT (X-SHOOTER)
}


\software{Molecfit \citep{Molecfit2015b}, Reflex \citep{Reflex}, astroquery \citep{astroquery}, galpy v1.6 \citep{Bovy2015}, lightkurve \citep{lk2018}, VOSA \citep{Bayo2008}, MOLLY (developed by Tom Marsh and available at
\url{http://deneb.astro.warwick.ac.uk/phsaap/software})}









\end{document}